\documentclass[a4paper,12pt,eqno]{article}
\usepackage{theorem}
\usepackage{latexsym,amssymb,amsfonts,amsmath}
\usepackage{graphicx}
\makeatletter
\@addtoreset{equation}{section}

\makeatother

\newtheorem{thm}{Theorem}[section]

\newtheorem{pro}[thm]{Proposition}

\theoremheaderfont{\scshape}
\newcommand{\RM}{\mathbb{R}}
\newcommand{\ZM}{\mathbb{Z}}

\newcommand{\CM}{\mathbb{C}}

\newcommand{\HM}{\mathbb{H}}

\newcommand{\ket}[1]{|#1\rangle}

\title{{\Large {\bf Quaternionic quantum walks}}}
\author{
{\small Norio Konno}\\
{\scriptsize Department of Applied Mathematics, 
Faculty of Engineering, 
Yokohama National University}\\
{\scriptsize Hodogaya, Yokohama 240-8501, Japan}\\
{\scriptsize e-mail: konno@ynu.ac.jp}\\
}
\date{\empty }
\begin{document}
\maketitle

\par\noindent
\begin{small}
\par\noindent
{\bf Abstract}. The discrete-time quantum walk (QW) has been extensively and intensively investigated for the last decade, whose coin operator is defined by a unitary matrix. We extend the QW to a walk determined by a unitary matrix whose component is quaternion. We call this model quaternionic quantum walk (QQW) and present some properties. This paper is the first step for the study on QQWs.  
\footnote[0]{
{\it Abbr. title:} Quaternionic quantum walks
}
\end{small}

\setcounter{equation}{0}
\section{Introduction \label{intro}}
The discrete-time quantum walk (QW) is a quantum version of the classical random walk and has been largely investigated for the last decade. The striking property of the QW is the spreading property of the walker. The standard deviation of the walker's position grows linearly in time, quadratically faster than classical random walk, i.e., ballistic spreading. On the other hand, a walker stays at the starting position: localization occurs. Interestingly, a quantum walker has both ballistic spreading and localization. The review and books on QWs are Kempe \cite{Kempe2003}, Kendon \cite{Kendon2007}, Venegas-Andraca \cite{VAndraca2008, Venegas2013}, Konno \cite{Konno2008b}, Cantero et al. \cite{CanteroEtAl2013}, Manouchehri and Wang \cite{MW2013}, Portugal \cite{P2013}. It is known that the quaternion was discovered by Hamilton in 1843. Quaternions can be considered as an extension of complex numbers. As for a survey on quaternions and matrices of quaternions, see Zhang \cite{Zhang1997}, for example. In this paper, we extend the QW to a walk given by a unitary matrix whose component is quaternion. Here we call the introduced walk {\it quaternionic quantum walk} (QQW). As for the detailed definition of the QQW, see Sect. \ref{def}. We explore a relation between QWs and QQWs in the present manuscript.

From now on we introduce some notations and a result on quaternions. Let $\RM$ be the set of real numbers. Let $\HM$ denote the set of quaternions of the form 
\begin{align*}
x=x_0+x_1i+x_2j+x_3k,
\end{align*}
where $x_0, x_1, x_2, x_3 \in \RM$ and 
\begin{align*}
i^2 &= j^2 = k^2 = -1, 
\\
ij & = -ji = k, \quad jk= -kj =i, \quad ki=-ki = j. 
\end{align*}
Then a direct computation gives 
\begin{pro}
For $x=x_0+x_1i+x_2j+x_3k \in \HM \> (x_0, x_1, x_2, x_3 \in \RM)$,
\begin{align*}
x^2 
=  x_0^2 - x_1^2 - x_2^2 - x_3^2 + 2 x_0 (x_1 i + x_2 j + x_3 k).
\end{align*}
\label{ajunko}
\end{pro}

For $x=x_0+x_1i+x_2j+x_3k \in \HM \> (x_0, x_1, x_2, x_3 \in \RM)$, let 
\begin{align*}
\overline{x} = x^{\ast} = x_0-x_1i-x_2j-x_3k
\end{align*}
be the conjugate of $x$, and 
\begin{align*}
|x| = \sqrt{x x^{\ast}} = \sqrt{x^{\ast} x} = \sqrt{x_0^2+x_1^2+x_2^2+x_3^2}
\end{align*}
be the modulus of $x$. Moreover we define $\Re x = x_0,$ the real part of $x$, and $\Im x = x_1i+x_2j+x_3k,$ the imaginary part of $x$.

Let $\mbox{\boldmath{M}}(n, \CM)$ and $\mbox{\boldmath{M}}(n, \HM)$ be the set of all $n \times n$ matrices with complex and quaternion entries, respectively. For $A=(a_{st}) \in \mbox{\boldmath{M}}(n, \HM)$, we put $\overline{A}= (\overline{a}_{st})=(a^{\ast}_{st})$ and $A^{\ast} = {}^T (\overline{A})$, where ${}^T A$ denotes the transpose of $A$. If $A A^{\ast} = I,$ then $A \in \mbox{\boldmath{M}}(n, \HM)$ is said to be unitary, where $I$ is the identity matrix. Let $\mbox{\boldmath{U}}(n, \CM)$ and $\mbox{\boldmath{U}}(n, \HM)$ denote the set of $n \times n$ unitary matrices with complex and quaternion entries, respectively.

The discrete-time QW on $\ZM$ with two chiralities is defined by $U \in \mbox{\boldmath{U}}(2, \CM)$, which was first intensively studied by Ambainis et al. \cite{AmbainisEtAl2001}, where $\ZM$ be the set of integers. Our QQW can be determined by $U \in \mbox{\boldmath{U}}(2, \HM)$. Let $H$ be the Hadamard gate, that is, 
\begin{align*}
H = \frac{1}{\sqrt{2}} 
\begin{bmatrix}
1 & 1 \\
1 & -1
\end{bmatrix}.
\end{align*}
If $U=H$, then the QQW becomes the {\it Hadamard walk} which has been well investigated in the study of QW.

The rest of the present paper is organized as follows. Section \ref{def} gives the detailed definition of QQWs on $\ZM$. In Sect. \ref{result}, we present some results on QQWs. Proofs of Theorems \ref{asanojunkodayo} and \ref{biwako1} are given in Sects. \ref{asanojunko} and \ref{unitary}, respectively. We consider stationary measures on QQWs for $a=0$ (Sect. \ref{azero}) and $b=0$ (Sect. \ref{bzero}), respectively. Section \ref{sum} is devoted to summary.

%%%%%%%%%%%%%%%%%%%%%%%%%%
\section{Model \label{def}}
The discrete-time QQW is a quaternion version of the QW with additional degree of freedom called chirality. The chirality takes values left and right, and it means the direction of the motion of the walker. At each time step, if the walker has the left chirality, it moves one step to the left, and if it has the right chirality, it moves one step to the right. Let us define
\begin{align*}
\ket{L} = 
\begin{bmatrix}
1 \\
0  
\end{bmatrix},
\qquad
\ket{R} = 
\begin{bmatrix}
0 \\
1  
\end{bmatrix},
\end{align*}
where $L$ and $R$ refer to the left and right chirality states, respectively.  

The walk is determined by $U \in \mbox{\boldmath{U}}(2, \HM)$, where   
\begin{align}
U =
\begin{bmatrix}
a & b \\
c & d
\end{bmatrix}.
\label{ohuro}
\end{align}
To define the dynamics of our model, we divide $U$ into two matrices:
\begin{eqnarray*}
P =
\begin{bmatrix}
a & b \\
0 & 0 
\end{bmatrix}, 
\quad
Q =
\begin{bmatrix}
0 & 0 \\
c & d 
\end{bmatrix},
\end{eqnarray*}
with $U =P+Q$. The important point is that $P$ (resp. $Q$) represents that the walker moves to the left (resp. right) at any position at each time step.

Let $\Psi_n (\in \HM^{\ZM})$ denote the state at time $n$ of the QQW on $\ZM$:  
\begin{align*}
\Psi_{n}
&= {}^T\![\cdots,\Psi_{n}^{L}(-1),\Psi_{n}^{R}(-1),\Psi_{n}^{L}(0),\Psi_{n}^{R}(0),\Psi_{n}^{L}(1),\Psi_{n}^{R}(1),\cdots ],
\\
&= {}^T\![\cdots, \Psi_{n}(-1), \Psi_{n}(0),\Psi_{n}(1), \cdots ],
\\
&= {}^T\!\left[\cdots,\begin{bmatrix}
\Psi_{n}^{L}(-1)\\
\Psi_{n}^{R}(-1)\end{bmatrix},\begin{bmatrix}
\Psi_{n}^{L}(0)\\
\Psi_{n}^{R}(0)\end{bmatrix},\begin{bmatrix}
\Psi_{n}^{L}(1)\\
\Psi_{n}^{R}(1)\end{bmatrix},\cdots\right],
\end{align*}
where $T$ means the transposed operation and $\Psi_{n}(x) = {}^T\![ \Psi_{n}^{L}(x), \Psi_{n}^{R}(x)]$ denotes the quaternion version of amplitude at time $n$ and position $x$. Then the time evolution of the walk is defined by 
\begin{align*}
\Psi_{n+1}(x)= P \Psi_{n} (x+1) +  Q \Psi_{n}(x-1).
\end{align*}
That is 
\begin{align*}
\begin{bmatrix}
\Psi_{n+1}^{L}(x)\\
\Psi_{n+1}^{R}(x)
\end{bmatrix}
=
\begin{bmatrix}           
a \Psi_{n}^{L}(x+1)+b \Psi_{n}^{R}(x+1)\\
c \Psi_{n}^{L}(x-1)+d \Psi_{n}^{R}(x-1)
\end{bmatrix}.
\end{align*}
Now let
\begin{align*}
U^{(s)}=\begin{bmatrix}
\ddots&\vdots&\vdots&\vdots&\vdots&\vdots&\cdots \\
\cdots&O&P&O&O&O&\cdots\\
\cdots&Q&O&P&O&O&\cdots\\
\cdots&O&Q&O&P&O&\cdots\\
\cdots&O&O&Q&O&P&\cdots\\
\cdots&O&O&O&Q&O&\cdots\\
\cdots&\vdots&\vdots&\vdots&\vdots&\vdots&\ddots
\end{bmatrix}\;\;\;
\mbox{with} \;\;\;
O=\begin{bmatrix}
0&0\\
0&0
\end{bmatrix}.
\end{align*}
Then the state of the QQW at time $n$ is given by
\begin{align}
\Psi_{n}=(U^{(s)})^{n}\Psi_{0},
\label{sankeien}
\end{align} 
for any $n\geq0$. Let $\mathbb{R}_{+}=[0,\infty)$. Here we introduce a map 
$\phi:(\mathbb{H}^{2})^{\mathbb{Z}}\rightarrow \mathbb{R}_{+}^{\mathbb{Z}}$
such that if
\begin{align*}
\Psi= {}^T\!\left[\cdots,\begin{bmatrix}
\Psi^{L}(-1)\\
\Psi^{R}(-1)\end{bmatrix},\begin{bmatrix}
\Psi^{L}(0)\\
\Psi^{R}(0)\end{bmatrix},\begin{bmatrix}
\Psi^{L}(1)\\
\Psi^{R}(1)\end{bmatrix},\cdots\right] \in(\mathbb{H}^{2})^{\mathbb{Z}},
\end{align*}
then 
\begin{align*}
\phi(\Psi) = {}^T\! 
\left[\ldots, 
|\Psi^{L}(-1)|^2 + |\Psi^{R}(-1)|^2, 
|\Psi^{L}(0)|^2 + |\Psi^{R}(0)|^2, 
|\Psi^{L}(1)|^2 + |\Psi^{R}(1)|^2, \ldots \right] 
\in \mathbb{R}_{+}^{\mathbb{Z}}.
\end{align*}
That is, for any $x \in \ZM$, 
\begin{align*}
\phi(\Psi) (x) = |\Psi^{L}(x)|^2 + |\Psi^{R}(x)|^2.
\end{align*}
Sometimes we identify $\phi(\Psi(x))$ with $\phi(\Psi) (x)$. Moreover we define the measure of the QQW at position $x$ by
\begin{align*}
\mu(x)=\phi(\Psi(x)) \quad (x \in \ZM).
\end{align*}
Now we are ready to introduce the set of stationary measures:  
\begin{align*}
{\cal M}_{s} 
&= {\cal M}_s (U)
\\
&= \left\{ \mu \in\mathbb{R}_{+}^{\mathbb{Z}} \setminus \{ \boldsymbol{0} \} : \mbox{there exists} \; \Psi_{0} \; \mbox{such that} \; \phi((U^{(s)})^{n}\Psi_{0})=\mu \; (n \ge 0) \right\},
\end{align*}
where $\boldsymbol{0}$ is the zero vector. We call the element of ${\cal M}_{s}$ the stationary measure of the QQW.

Next we consider the right (not left) eigenvalue problem of the QQW:
\begin{align}
U^{(s)} \Psi = \Psi \lambda \quad (\lambda \in \HM).
\label{samui}
\end{align}
Since the quaternions do not commute, it is necessary to treat $U^{(s)} \Psi = \lambda \Psi$ and $U^{(s)} \Psi = \Psi \lambda$ separately. Concerning left and right eigenvalues for the quaternionic matrix, and their properties, see Huang and So \cite{HuangSo2001}. From Eq. \eqref{samui}, we have
\begin{align*}
\left( U^{(s)} \right)^2 \Psi = U^{(s)} \left(U^{(s)} \Psi \right) =  \left( U^{(s)} \Psi \right) \lambda = \Psi \lambda^2.
\end{align*}
In general, we see
\begin{align*}
\left( U^{(s)} \right)^n \Psi = \Psi \lambda^n \quad (n \ge 1).
\end{align*}

We should remark that $|\lambda|=1$, since $U^{(s)}$ is unitary. We sometimes write $\Psi=\Psi^{(\lambda)}$ in order to emphasize the dependence on eigenvalue $\lambda$. Then we have $\phi (\Psi^{(\lambda)}) \in {\cal M}_s$.

We see that Eq. \eqref{samui} is equivalent to 
\begin{align}
\Psi^{L}(x) \lambda 
&= a \Psi^{L}(x+1) + b \Psi^{R} (x+1),
\label{yokoyama}
\\
\Psi^{R}(x) \lambda 
&= c \Psi^{L}(x-1) + d \Psi^{R}(x-1),
\label{taikan} 
\end{align}
for any $x \in \ZM$.

Put
\begin{align*}
\varphi = \begin{bmatrix} \alpha  \\ \beta  \end{bmatrix} \in \HM^2,
\end{align*}
with $\alpha, \beta \in \HM$ and $|\alpha|^2+|\beta|^2=1$. Let $\Psi_0 ^{\varphi}$ be the initial state for the QQW starting from $\varphi$ at the origin: 
\begin{align*}
\Psi_0 ^{\varphi} 
&={}^T \left[ \ldots, 
\begin{bmatrix} \Psi^{L} (-2) \\ \Psi^{R} (-2) \end{bmatrix}, 
\begin{bmatrix} \Psi^{L} (-1) \\ \Psi^{R} (-1) \end{bmatrix}, 
\begin{bmatrix} \Psi^{L} (0)  \\ \Psi^{R} (0)  \end{bmatrix}, 
\begin{bmatrix} \Psi^{L} (1)  \\ \Psi^{R} (1)  \end{bmatrix}, 
\begin{bmatrix} \Psi^{L} (2)  \\ \Psi^{R} (2)  \end{bmatrix}, 
\ldots \right],
\\
&= {}^T \left[ \ldots, 
\begin{bmatrix} 0 \\ 0 \end{bmatrix}, 
\begin{bmatrix} 0 \\ 0 \end{bmatrix}, 
\varphi, 
\begin{bmatrix} 0 \\ 0  \end{bmatrix}, 
\begin{bmatrix} 0 \\ 0  \end{bmatrix}, 
\ldots \right].
\end{align*}
The probability that quaternionic quantum walker at time $n$, $X_n= X_n ^{\varphi}$, with the initial $\Psi_0 ^{\varphi}$ exists at location $x \in \ZM$ is defined by
\begin{align*}
P \left( X_n = x \right) =  P \left( X_n ^{\varphi} = x \right) =  \phi \left( \left( U^{(s)} \right)^n \Psi_0 ^{\varphi} \right) (x).
\end{align*}

To compute $P \left( X_n = x \right)$, we consider the following quantity. For fixed $l$ and $m$ with $l+m=n$ and $-l+m=x$, we define  
\begin{align*}
\Xi_n (l,m)= \sum_{l_j,m_j}  P^{l_n}Q^{m_n}P^{l_{n-1}}Q^{m_{n-1}} \cdots P^{l_2}Q^{m_2} P^{l_1}Q^{m_1}
\end{align*}
summed over all $l_j, \> m_j \in \{0, 1\}$ satisfying $l_1+ \cdots +l_n=l, \> m_1+ \cdots + m_n = m,$ and $l_j + m_j =1 \> (1 \le j \le n)$. For example, $\Xi_3 (2,1) = P^2Q +PQP+ QP^2$. We put 
\begin{align*}
\varphi 
= \begin{bmatrix} \Psi^{L} (0)  \\ \Psi^{R} (0)  \end{bmatrix}
= \begin{bmatrix} \alpha  \\ \beta  \end{bmatrix} \in \HM^2.
\end{align*}
By definition, we see that
\begin{align*}
\Psi_n (x) = \Xi_n (l,m) \varphi,
\end{align*}
since $\Psi_n (x)$ is a two component vector of the quaternionic quantum walker being at position $x$ at time $n$ for initial state $\varphi$ at the origin, and $\Xi_n (l,m)$ is the sum of all possible paths in the trajectory consisting of $l$ steps left and $m$ steps right with $l=(n-x)/2$ and $m=(n+x)/2$.

From now on we consider $\Xi_n (l, m)$ for $n=3,4$ in the following. When $n=3$ case, we get
\begin{align*}
\Xi_3 (3, 0) 
&= 
\begin{bmatrix}
a^3 & a^2b \\
0   &   0  
\end{bmatrix}, \quad
\Xi_3 (2, 1) = 
\begin{bmatrix}
abc+bca & abd+bcb \\
ca^2 & cab  
\end{bmatrix}, 
\\
\Xi_3 (1, 2) 
&= 
\begin{bmatrix}
bdc & bd^2 \\
cbc+dca & cbd+dcb
\end{bmatrix}, \quad
\Xi_3 (0, 3) = 
\begin{bmatrix}
0 & 0 \\
d^2c & d^3
\end{bmatrix}. 
\end{align*}

When $n=4$ case, we obtain
\begin{align*}
\Xi_4 (4, 0) 
&= 
\begin{bmatrix}
a^4 & a^3b \\
0   &   0  
\end{bmatrix}, \quad
\Xi_4 (3, 1) = 
\begin{bmatrix}
abca+bca^2+a^2bc & a^2bd+abcb+bcab \\
ca^3 & ca^2b  
\end{bmatrix}, 
\\
\Xi_4 (2, 2) 
&= 
\begin{bmatrix}
bdca+abdc+bcbc & abd^2+bdcb+bcbd \\
cbca+dca^2+cabc & dcab+cbcb+cabd
\end{bmatrix}, 
\\
\Xi_4 (1, 3) 
&= 
\begin{bmatrix}
bd^2c & bd^3 \\
dcbc+d^2ca+cbdc & dcbd+d^2cb+cbd^2
\end{bmatrix},
\quad
\Xi_4 (0, 4) 
= 
\begin{bmatrix}
0 & 0 \\
d^3c   &   d^4  
\end{bmatrix}.
\end{align*}

As an example, we deal with the following QQW defined by 
\begin{align*}
U =
\frac{1}{\sqrt{2}}
\begin{bmatrix}
1 & i \\
j & k
\end{bmatrix}.
\end{align*}
When $n=3$,
\begin{align*}
\Xi_3 (3, 0)
&= 
\frac{1}{2\sqrt{2}}
\begin{bmatrix}
1 & i \\
0 & 0  
\end{bmatrix}, \quad
\Xi_3 (2, 1) = 
\frac{1}{2\sqrt{2}}
\begin{bmatrix}
2k & 0 \\
j & -k  
\end{bmatrix}, 
\\
\Xi_3 (1,2) 
&= 
\frac{1}{2\sqrt{2}}
\begin{bmatrix}
1 & -i \\
0 & 2
\end{bmatrix}, \quad 
\Xi_3 (0,3) 
= 
\frac{1}{2\sqrt{2}}
\begin{bmatrix}
0 & 0 \\
-j & -k
\end{bmatrix}.
\end{align*}
When $n=4$,
\begin{align*}
\Xi_4 (4, 0)
&= 
\frac{1}{4}
\begin{bmatrix}
1 & i \\
0   &   0  
\end{bmatrix}, \quad
\Xi_4 (3, 1) = 
\frac{1}{4}
\begin{bmatrix}
3k & j \\
j & -k  
\end{bmatrix}, \quad
\Xi_4 (2, 2) 
= 
\frac{1}{4}
\begin{bmatrix}
1 & i \\
i & 1
\end{bmatrix}, 
\\
\Xi_4 (1, 3) 
&= 
\frac{1}{4}
\begin{bmatrix}
-k & j \\
j & 3k
\end{bmatrix},
\quad
\Xi_4 (0, 4) 
= 
\frac{1}{4}
\begin{bmatrix}
0 & 0 \\
i & 1  
\end{bmatrix}.
\end{align*}
If we take an initial state at the origin $\varphi = {}^T[1/\sqrt{2}, j/\sqrt{2}]$, then we have
\begin{align*}
\Xi_3 (3, 0) \varphi
&= 
\frac{1}{4}
\begin{bmatrix}
1+k \\
0  
\end{bmatrix}, \quad
\Xi_3 (2, 1) \varphi = 
\frac{1}{4}
\begin{bmatrix}
2k \\
i+j  
\end{bmatrix}, 
\\
\Xi_3 (1,2) \varphi
&= 
\frac{1}{4}
\begin{bmatrix}
1-k \\
2j
\end{bmatrix}, \quad 
\Xi_3 (0,3) \varphi
= 
\frac{1}{4}
\begin{bmatrix}
0 \\
i-j
\end{bmatrix},
\end{align*}
\begin{align*}
\Xi_4 (4, 0) \varphi 
&= \frac{1}{4 \sqrt{2}}
\begin{bmatrix}
1+k \\
0  
\end{bmatrix}, \quad
\Xi_4 (3, 1) \varphi 
= \frac{1}{4 \sqrt{2}}
\begin{bmatrix}
-1+3k \\
i+j  
\end{bmatrix}, \quad
\Xi_4 (2, 2) \varphi 
= \frac{1}{4 \sqrt{2}}
\begin{bmatrix}
1+k \\
i+j  
\end{bmatrix}, 
\\
\Xi_4 (1, 3) \varphi 
&= \frac{1}{4 \sqrt{2}}
\begin{bmatrix}
-1-k \\
-3i+j
\end{bmatrix}, \quad
\Xi_4 (0, 4) \varphi 
= \frac{1}{4 \sqrt{2}}
\begin{bmatrix}
0 \\
i+j
\end{bmatrix}.
\end{align*}
Therefore we get
\begin{align*}
P(X_3  = -3) &=  P(X_3  = 3) =1/8, \quad P(X_3  = -1) = P(X_3  = 1)=3/8, 
\\
P(X_4  = -4) &=  P(X_4  = 4) =1/16, \\
P(X_4  = -2) &= P(X_4  = 2)=6/16, \quad P(X_4  = 0) = 2/16. 
\end{align*}
In fact, it is noted that the probability distributions for $n=0,1,2,3,4$ are the same as those of the symmetric Hadamard walk with initial state at the origin, e.g., $\varphi = {}^T[1/\sqrt{2}, i/\sqrt{2}]$.

From now on, we treat general $\Xi_n (l,m)$. For example, 
\begin{align*}
\Xi_4 (3,1) = QP^3 + PQP^2 + P^2QP + P^3Q.
\end{align*}
Here we find a nice relation: $P^2 = aP.$ We introduce the following $2 \times 2$ matrices, $R$ and $S$:
\begin{align*}
R=
\left[
\begin{array}{cc}
c & d \\
0 & 0 
\end{array}
\right], 
\quad
S=
\left[
\begin{array}{cc}
0 & 0 \\
a & b 
\end{array}
\right].
\end{align*}
Then we obtain the next table of products of matrices, $P, \> Q, \> R,$ and $S$:
\par
\
\par
\begin{center}
\begin{tabular}{c|cccc}
  & $P$ & $Q$ & $R$ & $S$  \\ \hline
$P$ & $aP$ & $bR$ & $aR$ & $bP$  \\
$Q$ & $cS$ & $dQ$& $cQ$ & $dS$ \\
$R$ & $cP$ & $dR$& $cR$ & $dP$ \\
$S$ & $aS$ & $bQ$ & $aQ$ & $bS$ 
\label{pqrs}
\end{tabular}
\end{center}
%\begin{center}
%Table 1.1
%\end{center}
\par
\
\par\noindent
where $PQ=bR$, for example. By using this table, we have

\begin{align*}
\overbrace{PP \cdots P}^{w_1} \overbrace{QQ \cdots Q}^{w_2} \overbrace{PP \cdots P}^{w_3} 
&
\cdots \overbrace{QQ \cdots Q}^{w_{2 \gamma}} \overbrace{PP \cdots P}^{w_{2 \gamma+1}}
\\
&= a^{w_1-1} b d^{w_2-1} c a^{w_3-1} b \cdots d^{w_{2 \gamma}-1} c a^{w_{2 \gamma+1}-1} P, 
\\
\overbrace{QQ \cdots Q}^{w_1} \overbrace{PP \cdots P}^{w_2} \overbrace{QQ \cdots Q}^{w_3} 
&
\cdots \overbrace{PP \cdots P}^{w_{2 \gamma}} \overbrace{QQ \cdots Q}^{w_{2 \gamma+1}}
\\
&= d^{w_1-1} c a^{w_2-1} b d^{w_3-1} c \cdots a^{w_{2 \gamma}-1} b d^{w_{2 \gamma+1}-1} Q, 
\end{align*}
where $w_1, w_2, \ldots , w_{2 \gamma+1} \ge 1$ and $\gamma \ge 1$.

We should remark that $P, \> Q, \> R,$ and $S$ form an orthogonal basis of the vector space of $2 \times 2$ quaternionic matrices with respect to the trace inner product $\langle A | B \rangle = $ tr$(A^{\ast}B)$. So $\Xi_n (l,m)$ has the following form: 
\begin{align*}
\Xi_n (l,m) = p_n (l,m) P + q_n (l,m) Q + r_n (l,m) R + s_n (l,m) S.
\end{align*}
Next problem is to obtain explicit forms of $p_n (l,m), q_n (l,m), r_n (l,m)$, and $s_n (l,m)$. In the case of $n=l+m=4$ with $l=3,\> m=1$, we have $\Xi_4 (3,1) = (abc+bca) P + a^2b R + c a^2 S$. So this case is 
\begin{align*}
p_4 (3,1)=abc+bca, \quad q_4 (3,1)=0, \quad r_4 (3,1)=a^2b, \quad s_4 (3,1)=ca^2.
\end{align*}
If $a,b,c,d \in \CM$ (QW case), then we have
\begin{align*}
p_4 (3,1)=2abc, \quad q_4 (3,1)=0, \quad r_4 (3,1)=a^2b, \quad s_4 (3,1)=a^2c.
\end{align*}
However it would be hard to obtain an explicit form $\Xi_n (l,m)$ like that of QW case (see Lemma 1 in \cite{Konno2002}, or Lemma 2 on \cite{Konno2005}, for example).

%%%%%%%%%%%%%%%%%%%%%%%%%%%%%%%%%%%%%%%
\section{Results \label{result}}
In this section, we present our results on QQWs. Let 
\begin{align*}
\Psi (X) = \left\{ \varphi = 
\begin{bmatrix} \alpha  \\ \beta  \end{bmatrix} \in X^2
: |\alpha|^2 + |\beta|^2 =1  \right\},
\end{align*}
for $X = \RM, \CM, \HM$. Here we introduce the following set of measures. For any fixed $U \in \mbox{\boldmath{U}} (2,X)$ with $X = \RM, \CM, \HM$, 
\begin{align*}
{\cal M}^{loc}_n (U;(\varphi,Y))
&= \left\{ \phi \left( (U^{(s)})^{n} \Psi_{0} ^{\varphi} \right) : \varphi \in \Psi (Y) \right\} \quad (Y = \RM, \CM, \HM),
\\
{\cal M}^{glob}_n (U;(\Psi_0,Y))
&= \left\{ \phi \left( (U^{(s)})^{n} \Psi_{0} \right) : \Psi_{0} \in Y^{\ZM} \setminus \{ \boldsymbol{0} \} \right\} \quad (Y = \RM, \CM, \HM),
\end{align*}
where $n=0,1,2, \ldots$. By definition, for any $U \in \mbox{\boldmath{U}} (2,X)$ with $X = \RM, \CM, \HM$, we have 
\begin{align*}
{\cal M}^{loc}_0 (U;(\varphi,Y)) = \left\{ \delta_0 \right\}, 
\end{align*}
where $Y = \RM, \CM, \HM$. Here $\delta_x$ is the delta measure at position $x \in \ZM$. It is trivial that for any fixed $U \in \mbox{\boldmath{U}} (2,X)$ with $X = \RM, \CM, \HM$, 
\begin{align*}
{\cal M}^{loc}_n (U;(\varphi,\RM)) \subseteq {\cal M}^{loc}_n (U;(\varphi,\CM)) \subseteq {\cal M}^{loc}_n (U;(\varphi,\HM)) \quad (n \ge 0),
\end{align*}
and
\begin{align*}
{\cal M}^{glob}_n (U;(\Psi_{0},\RM)) \subseteq {\cal M}^{glob}_n (U;(\Psi_{0},\CM)) \subseteq {\cal M}^{glob}_n (U;(\Psi_{0},\HM)) \quad (n \ge 0).
\end{align*}
In this setting, we obtain
%%%%%%%%%%%%%%%%%%%%%%%%%%%%%%%%%%%%%%%%%%%%%%%%%%%%%%%
\begin{thm}
For any $U \in \mbox{\boldmath{U}} (2,\RM)$, 
\begin{align*}
{\cal M}^{loc}_n (U;(\varphi,\CM)) = {\cal M}^{loc}_n (U;(\varphi,\HM)) \quad (n \ge 0).
\end{align*}
\label{asanojunkodayo}
\end{thm}
%%%%%%%%%%%%%%%%%%%%%%%%%%%%%%%%%%%%%%%%%%%%%%%%%%%%%%%
The typical example is the Hadamard walk given by $U=H \in \mbox{\boldmath{U}} (2,\RM)$. The proof will appear in Section \ref{asanojunko}. 

Next we consider a relation between ${\cal M}^{loc}_n (U;(\varphi,\RM))$ and ${\cal M}^{loc}_n (U;(\varphi,\CM))$ for the Hadamard walk, that is, $U=H$. In this case, we have
\begin{align*}
P \left( X_1 = -1 \right) = \frac{1}{2} | \alpha + \beta |^2, \quad P \left( X_1 = 1 \right) = \frac{1}{2} | \alpha - \beta |^2.
\end{align*}
Then we see that $P (X_1 = -1) = P (X_1 = 1)$ if and only if $\Re (\alpha \overline{\beta}) = 0$, where $\Re (x)$ is the real part of $x \in \HM$. In a similar fashion, we obtain  
\begin{align*}
P \left( X_2 = -2 \right) = \frac{1}{4} | \alpha + \beta |^2, \quad P \left( X_2 = 2 \right) = \frac{1}{4} | \alpha - \beta |^2.
\end{align*}
Then we see that $P (X_2 = -2) = P (X_2 = 2)$ if and only if $\Re (\alpha \overline{\beta}) = 0$. Here we introduce the set of symmetric measures:
\begin{align*}
{\cal M}_{sym} 
= \left\{ \mu \in\mathbb{R}_{+}^{\mathbb{Z}} \setminus \{ \boldsymbol{0} \} : \mu (x) = \mu (-x) \>\> (x \in \ZM) \right\}.
\end{align*}Therefore we have the following result:
%%%%%%%%%%%%%%%%%%%%%%%%%%%%%%%%%%%%%%%%%%%%%%%%%%%%%%%
\begin{pro}
For any $Y = \RM, \CM, \HM$, we obtain
\begin{align*}
{\cal M}^{loc}_1 (H;(\varphi,Y)) \cap {\cal M}_{sym} 
&= \left\{ \frac{1}{2} \delta_{-1} + \frac{1}{2} \delta_{1} \right\}, 
\\
{\cal M}^{loc}_2 (H;(\varphi,Y)) \cap {\cal M}_{sym}
&= \left\{ \frac{1}{4} \delta_{-2} + \frac{2}{4} \delta_{0} + \frac{1}{4} \delta_{2} \right\}.
\end{align*}
\label{asjunko}
\end{pro}
%%%%%%%%%%%%%%%%%%%%%%%%%%%%%%%%%%%%%%%%%%%%%%%%%%%%%%%
Furthermore, we have
\begin{align}
P \left( X_3 = -3 \right) 
&= \frac{1}{8} |\alpha + \beta|^2, \quad P \left( X_3 = 3 \right) = \frac{1}{8} |\alpha - \beta|^2,
\label{aji1}
\\
P \left( X_3 = -1 \right) 
&= \frac{1}{8} \left\{ 4 |\alpha|^2 + |\alpha + \beta|^2 \right\}, \quad P \left( X_3 = 1 \right) = \frac{1}{8} \left\{ 4|\beta|^2 + |\alpha - \beta|^2 \right\}.
\label{aji2}
\end{align}
Then we see that ``$P (X_3 = -3) = P (X_3 = 3)$ and $P(X_3 = -1) = P(X_3 = 1)$" if and only if ``$\Re (\alpha \overline{\beta}) = 0$ and $|\alpha|=|\beta|=1/\sqrt{2}$". Therefore we have 
%%%%%%%%%%%%%%%%%%%%%%%%%%%%%%%%%%%%%%%%%%%%%%%%%%%%%%%
\begin{pro}
\begin{align*}
{\cal M}^{loc}_3 (H;(\varphi,\RM)) \cap {\cal M}_{sym}  
&= \emptyset, 
\\
{\cal M}^{loc}_3 (H;(\varphi,Y)) \cap {\cal M}_{sym}
&= \left\{ \frac{1}{8} \delta_{-3} + \frac{3}{8} \delta_{-1} + \frac{3}{8} \delta_{1} + \frac{1}{8} \delta_{3} \right\},
\end{align*}
for $Y = \CM, \HM$.
\label{asajunko}
\end{pro}
%%%%%%%%%%%%%%%%%%%%%%%%%%%%%%%%%%%%%%%%%%%%%%%%%%%%%%%
Thus we see that
\begin{align*}
{\cal M}^{loc}_3 (H;(\varphi,\RM)) \subset {\cal M}^{loc}_3 (H;(\varphi,\CM)).
\end{align*}
So, compared with Theorem \ref{asanojunkodayo}, the following does not hold; for any $U \in \mbox{\boldmath{U}} (2,\RM)$, 
\begin{align*}
{\cal M}^{loc}_n (U;(\varphi,\RM)) = {\cal M}^{loc}_n (U;(\varphi,\CM)) \quad (n \ge 0).
\end{align*}

As in the similar way of our previous paper \cite{KonnoTakei2014}, we obtain the following results; Theorems \ref{biwako1}, \ref{biwako2}, and \ref{biwako3}. Remark that for any $U \in \mbox{\boldmath{U}} (2,\HM)$, the unitarity of $U$ implies that it is enough to consider three cases: $abcd \not=0, \> a=0,$ and $b=0$. For any $c>0$, $\mu_{u}^{(c)}$ denotes the uniform measure with parameter $c$, i.e., 
\begin{align*}
\mu_{u}^{(c)} (x) = c \qquad (x \in \ZM). 
\end{align*}
Let ${\cal M}_{unif} = \{ \mu_{u}^{(c)} : c>0 \}$ be the set of uniform measures on $\ZM$. 
\begin{thm}
\label{biwako1}
For any $U \in \mbox{\boldmath{U}} (2, \HM)$, we have
\begin{align}
{\cal M}_{unif} \subseteq {\cal M}_s.
\end{align}
\end{thm}
Let ${\cal M}_{exp}$ be the set of the measures $\mu$ having exponential decay with respect to the position, i.e., $\mu$ satisfies that there exist positive constants $C_+, C_0, C_-$, and $\gamma \in (0,1)$ such that 
\begin{align*}
\mu (x) = 
\left\{ 
\begin{array}{cc}
C_+ \gamma^{-|x|} & (x \ge 1), \\
C_0 & (x =0), \\
C_- \gamma^{-|x|} & (x \le -1). 
\end{array}
\right.
\end{align*}
Furthermore we obtain the following result for $a=0$ case.
\begin{thm}
\label{biwako2}
For any $U \in \mbox{\boldmath{U}} (2, \HM)$ with $a=0$, we see 
\begin{align*}
{\cal M}_s \setminus \left({\cal M}_{unif} \cup {\cal M}_{exp} \right) \not= \emptyset.
\end{align*}
\end{thm}
The proof will be given in Sect. \ref{azero}. For $b=0$ case, we show 
\begin{thm}
\label{biwako3}
For any $U \in \mbox{\boldmath{U}} (2, \HM)$ with $b=0$, we see 
\begin{align*}
{\cal M}_s = {\cal M}_{unif}. 
\end{align*}
\end{thm}
Concerning the proof, see Sect. \ref{bzero}. For the rest ($abcd \not= 0$ case), we do not have any corresponding interesting results on QQWs at the present stage.

%%%%%%%%%%%%%%%%%%%%%%%%%%%%%%%%
\section{Proof of Theorem \ref{asanojunkodayo} \label{asanojunko}}
By definition, it is obvious that 
\begin{align*}
{\cal M}^{loc}_n (U;(\varphi,\CM)) \subset {\cal M}^{loc}_n (U;(\varphi,\HM)) \quad (n \ge 0).
\end{align*}
Thus it is enough to show 
\begin{align*}
{\cal M}^{loc}_n (U;(\varphi,\HM)) \subset {\cal M}^{loc}_n (U;(\varphi,\CM)) \quad (n \ge 0).
\end{align*}
That is, for any $\varphi = {}^T [\alpha, \beta] \in \HM^2$ with $|\alpha|^2+|\beta|^2=1$, there exists $\widetilde{\varphi} = {}^T [\widetilde{\alpha}, \widetilde{\beta}] \in \CM^2$ with $|\widetilde{\alpha}|^2+|\widetilde{\beta}|^2=1$ such that 
\begin{align}
P \left( X_n ^{\varphi} = x \right) = P \left( X_n ^{\widetilde{\varphi}} = x \right) 
\end{align}
for any $n=0,1,2, \ldots$ and $x \in \ZM.$

First we see that $\alpha, \beta \in \HM$ with $|\alpha|^2+|\beta|^2=1$ can be written as
\begin{align}
\alpha 
&= \left\{ \cos \theta_{\alpha} + \left( \alpha_x i + \alpha_y j + \alpha_z k \right) \sin \theta_{\alpha} \right\} \cos \xi,
\label{ikuko1}
\\
\beta 
&= \left\{ \cos \theta_{\beta} + \left( \beta_x i + \beta_y j + \beta_z k \right) \sin \theta_{\beta} \right\} \sin \xi,
\label{ikuko2}
\end{align}
where $\theta_{\alpha}, \> \theta_{\beta} \in [0, 2\pi), \> \xi \in [0, \pi/2], \> \alpha_x, \alpha_y, \alpha_z, \beta_x, \beta_y, \beta_z \in \RM$ with 
\begin{align*}
\alpha_x ^2 + \alpha_y ^2 + \alpha_z ^2 = \beta_x ^2 + \beta_y ^2 + \beta_z ^2 = 1.
\end{align*}
From $U \in \mbox{\boldmath{U}} (2,\RM)$ and Lemma 1 in \cite{Konno2002} (or Lemma 2 in \cite{Konno2005}), we have
\begin{align*}
\Xi_n (l,m) = 
\begin{bmatrix}
r_{11} & r_{12} \\
r_{21} & r_{22}  
\end{bmatrix},
\end{align*}
where $r_{ab} \in \RM \> (a,b \in \{1,2\})$. Then we obtain 
\begin{align}
P \left( X_n ^{\varphi} = x \right) = || \Xi_n (l,m) \varphi ||^2 = A |\alpha|^2 + B |\beta|^2 + C \Re \left( \alpha \overline{\beta} \right), 
\label{ikuko3}
\end{align}
where $n=l+m, \> x=-l+m$ and
\begin{align*}
A = r_{11}^2 + r_{21}^2, \quad B = r_{12}^2 + r_{22}^2, \quad C = 2 \left( r_{11} r_{12} + r_{21} r_{22} \right).
\end{align*}
Combining Eqs. \eqref{ikuko1} and \eqref{ikuko2} with Eq. \eqref{ikuko3} implies 
\begin{align}
P \left( X_n ^{\varphi} = x \right) = A \cos^2 \xi + B \sin^2 \xi + C \left( \cos \theta_{\alpha} \cos \theta_{\beta} + \gamma \sin \theta_{\alpha} \sin \theta_{\beta} \right) \cos \xi \sin \xi, 
\label{ikuko4}
\end{align}
where $\gamma = \alpha_x \beta_x + \alpha_y \beta_y + \alpha_z \beta_z.$ If we take $\widetilde{\alpha}, \widetilde{\beta} \in \CM$ with 
\begin{align*}
\widetilde{\alpha} 
= \left( \cos \widetilde{\theta}_{\alpha} + i \sin \widetilde{\theta}_{\alpha} \right) \cos \widetilde{\xi}, \quad 
\widetilde{\beta} 
= \left( \cos \widetilde{\theta}_{\beta} + i \sin \widetilde{\theta}_{\beta} \right) \sin \widetilde{\xi}
\end{align*}
where $\widetilde{\theta}_{\alpha}, \> \widetilde{\theta}_{\beta} \in [0, 2\pi), \> \widetilde{\xi} \in [0, \pi/2]$, then we have 
\begin{align}
P \left( X_n ^{\widetilde{\varphi}} = x \right) = A \cos^2 \widetilde{\xi} + B \sin^2 \widetilde{\xi} + C \cos \left( \widetilde{\theta}_{\alpha} - \widetilde{\theta}_{\beta} \right) \cos \widetilde{\xi} \sin \widetilde{\xi}. 
\label{ikuko5}
\end{align}
We should remark that $|\gamma| \le 1$. So we see 
\begin{align*}
\left| \cos \theta_{\alpha} \cos \theta_{\beta} + \gamma \sin \theta_{\alpha} \sin \theta_{\beta} \right| \le 1.
\end{align*}
Therefore if we choose $\widetilde{\xi} = \xi$ and $\widetilde{\theta}_{\alpha}, \> \widetilde{\theta}_{\beta}$ satisfying
\begin{align*}
\cos \left( \widetilde{\theta}_{\alpha} - \widetilde{\theta}_{\beta} \right) = \cos \theta_{\alpha} \cos \theta_{\beta} + \gamma \sin \theta_{\alpha} \sin \theta_{\beta}, 
\end{align*}
then Eqs. \eqref{ikuko4} and \eqref{ikuko5} give 
\begin{align*}
P \left( X_n ^{\varphi} = x \right) = P \left( X_n ^{\widetilde{\varphi}} = x \right)
\end{align*}
for any $n=0,1,2, \ldots$ and $x \in \ZM$. Thus the proof is completed.

%%%%%%%%%%%%%%%%%%%%%%%%%%%%%%%%%%%%%%%%
\section{Proof of Theorem \ref{biwako1} \label{unitary}}
This section gives a proof of Theorem \ref{biwako1}, i.e., ${\cal M}_{unif} \subseteq {\cal M}_s (U)$ for any $U \in \mbox{\boldmath{U}} (2, \HM).$ This proof is the same as that in \cite{KonnoTakei2014}. So we omit the details. First we consider the following initial state: for any $x \in \ZM$, 
\begin{align*}
\Psi_{0} (x) = \varphi = \begin{bmatrix} \alpha  \\ \beta  \end{bmatrix}\in \HM^2,
\end{align*}
where $||\varphi||^2 = |\alpha|^2+|\beta|^2>0$. Remark that $\Psi_{0} (x)$ does not depend on the position $x$. Then we have 
\begin{align*}
\Psi_{1} (x) = P \Psi_{0} (x+1) +  Q \Psi_{0}(x-1) = (P+Q) \varphi = U \varphi.
\end{align*}
In a similar fashion, we get $\Psi_{n} (x) = U^n \varphi$ for any $n =0,1,2, \ldots$ and $x \in \ZM$. Thus we have 
\begin{align*}
\mu_n (x) = || \Psi_{n} (x) ||^2 =|| U^n \varphi ||^2 = ||\varphi ||^2 (= |\alpha|^2 + |\beta|^2),
\end{align*} 
since $U$ is unitary. That is, this measure $\mu_0$ satisfies $\mu_0 = \mu_{u}^{(c)}$ with $c=||\varphi ||^2$ and $\mu_n (x)= \mu_0 (x) \> (n \ge 1, \> x \in \ZM).$ Therefore the proof is completed.

It is noted that we can easily generalize Theorem \ref{biwako1} for an $N$-state QQW on $\ZM$ determined by the $N \times N$ unitary matrix, $U \in \mbox{\boldmath{U}} (N, \HM)$ like QW case (see \cite{KonnoTakei2014}).

%%%%%%%%%%%%%%%%%%%%%%%%%%%%%%%%%%%%%%%%%%%%%%%%%%%%%
%\section{Case $abcd \not=0$ \label{abcdzero}}

%%%%%%%%%%%%%%%%%%%%%%%%%%%%%%%%%%%%%%%
\section{Case $a=0$ \label{azero}}
This section deals with right eigenvalue problem and stationary measures on QQWs for $a=0$. In this case, $U$ can be expressed as 
\begin{align*}
U=
\begin{bmatrix}
0 & b  \\
c & 0
\end{bmatrix},
\end{align*}
where $b, c \in \HM$ and $|b|=|c|=1$.

First we consider
\begin{align*}
U=
\begin{bmatrix}
0 & 1  \\
1 & 0
\end{bmatrix}.
\end{align*}
From Eqs. \eqref{yokoyama} and \eqref{taikan}, we see that for any $x \in \ZM$,
\begin{align*}
\Psi^L(x) (\lambda^2 - 1) = 0.
\end{align*}
From Proposition \ref{ajunko}, we get two eigenvalues $\lambda_{\pm} = \pm 1$. As an initial state, we consider $\Psi^{(\pm)}$ corresponding to $\lambda_{\pm}$ as follows;
\begin{align}
\Psi^{(\pm)} = {}^T \left[ \ldots, 
\begin{bmatrix} \Psi^{(\pm,L)} (-2) \\ \Psi^{(\pm,R)} (-2) \end{bmatrix}, 
\begin{bmatrix} \Psi^{(\pm,L)} (-1) \\ \Psi^{(\pm,R)} (-1) \end{bmatrix}, 
\begin{bmatrix} \Psi^{(\pm,L)} (0)  \\ \Psi^{(\pm,R)} (0)  \end{bmatrix}, 
\begin{bmatrix} \Psi^{(\pm,L)} (1)  \\ \Psi^{(\pm,R)} (1)  \end{bmatrix}, 
\begin{bmatrix} \Psi^{(\pm,L)} (2)  \\ \Psi^{(\pm,R)} (2)  \end{bmatrix}, 
\ldots \right].
\label{huyumiHayaku}
\end{align}
Here for any $x \in \ZM$, 
\begin{align}
\Psi^{(\pm,L)} (2x) &= \alpha_{2x}, \>\> \Psi^{(\pm,R)} (2x) = \beta_{2x}, 
\nonumber
\\
\Psi^{(\pm,L)} (2x-1) &= \beta_{2x} \lambda_{\pm}, \>\> \Psi^{(\pm,R)} (2x+1) = \alpha_{2x} \lambda_{\pm} ,
\label{huyumiHayaku2}
\end{align}
where $\alpha_{2x}, \> \beta_{2x} \in \HM$ with $\alpha_{2x} \beta_{2x} \not= 0$. In fact, we have $U^{(s)} \Psi^{(\pm)} = \Psi^{(\pm)} \lambda_{\pm}.$ Therefore
\begin{align}
(U^{(s)})^n \Psi^{(\pm)} = \Psi^{(\pm)} \lambda_{\pm}^n.
\label{huyumiHayaku3}
\end{align}

Let $\mu_n ^{(\Psi^{(\pm)})} = \phi ((U^{(s)})^n \Psi^{(\pm)})$ and  
\begin{align*}
\mu_n ^{(\Psi^{(\pm)})}
= {}^T \left[ \ldots, \mu_n^{(\Psi^{(\pm)})} (-2), \mu_n^{(\Psi^{(\pm)})} (-1), \mu_n^{(\Psi^{(\pm)})} (0), \mu_n^{(\Psi^{(\pm)})} (1), \mu_n^{(\Psi^{(\pm)})} (2), \ldots \right].
\end{align*}
From Eqs. \eqref{huyumiHayaku}, \eqref{huyumiHayaku2}, and \eqref{huyumiHayaku3}, we obtain  
\begin{align*}
\mu_n ^{(\Psi^{(\pm)})} = {}^T \left[ \ldots, |\alpha_{-2}|^2+|\beta_{-2}|^2, |\alpha_{-2}|^2+|\beta_0|^2, |\alpha_0|^2+|\beta_0|^2, |\alpha_0|^2+|\beta_2|^2, |\alpha_2|^2+|\beta_2|^2, \ldots \right].
\end{align*}
Therefore we see that for any $n \ge 0$, $\mu_n ^{(\Psi^{(\pm)})} = \mu_0 ^{(\Psi^{(\pm)})}$. So $\mu_0 ^{(\Psi^{(\pm)})}$ becomes a stationary measure, that is, $\mu_0 ^{(\Psi^{(\pm)})} \in {\cal M}_s (U)$. Moreover $\mu_n ^{(\Psi^{(\pm)})} (2x) = |\alpha_{2x}|^2+|\beta_{2x}|^2 \> (x \in \ZM)$. So in general, stationary measure $\mu_0^{(\Psi^{(\pm)})}$ is a non-uniform and non-exponential decay measure. Therefore we obtain
\begin{align*}
{\cal M}_s (U) \setminus \left( {\cal M}_{unif} \cup {\cal M}_{exp} \right) \not= \emptyset.
\end{align*}

Next we consider the following case:
\begin{align*}
U=
\begin{bmatrix}
0 & 1  \\
-1 & 0
\end{bmatrix}.
\end{align*}
By Eqs. \eqref{yokoyama} and \eqref{taikan}, we see that for any $x \in \ZM$,
\begin{align*}
\Psi^L(x) (\lambda^2 + 1) = 0.
\end{align*}

By Proposition \ref{ajunko}, we have infinite eigenvalues: 
\begin{align*}
\lambda = x_1 i + x_2 j + x_3 k \quad (x_1^2+x_2^2+x_3^2=1). 
\end{align*}
As an initial state, we consider $\Psi^{(\lambda)}$ corresponding to $\lambda$ with 
\begin{align*}
\Psi^{(\lambda,L)} (2x) &= \alpha_{2x}, \>\> \Psi^{(\lambda,R)} (2x) = \beta_{2x}, 
\nonumber
\\
\Psi^{(\lambda,L)} (2x-1) &= -\beta_{2x} \lambda, \>\> \Psi^{(\lambda,R)} (2x+1) = \alpha_{2x} \lambda,
%\label{huyumiHayakuQ2}
\end{align*}
where $\alpha_{2x}, \> \beta_{2x} \in \HM$ with $\alpha_{2x} \beta_{2x} \not= 0$. As in the previous case, we obtain the same conclusion: 
\begin{align*}
{\cal M}_s (U) \setminus \left( {\cal M}_{unif} \cup {\cal M}_{exp} \right) \not= \emptyset.
\end{align*}
Similarly, we can extend this result to the general case $b, c \in \HM$ with $|b|=|c|=1$ and $a=d=0$.

%%%%%%%%%%%%%%%%%%%%%%%%%%%%%%%%%%%%%%
\section{Case $b=0$ \label{bzero}}
This section is devoted to stationary measures on QQWs for $b=0$. For this case, we see that $U$ can be written as 
\[
U=
\begin{bmatrix}
a & 0 \\
0 & d 
\end{bmatrix},
\]
where $a, d \in \HM$ with $|a|=|d|=1$. Here we introduce the following set of measures:
\begin{align*}
{\cal M}_{n} 
&= {\cal M}_n (U)
\\
&= \left\{ \mu \in\mathbb{R}_{+}^{\mathbb{Z}} \setminus \{ \boldsymbol{0} \} : \mbox{there exists} \; \Psi_{0} \; \mbox{such that} \right. 
\\
& \qquad \qquad \qquad \qquad \qquad \left.
\mbox{for any $n \ge 0$,} \> \>  \phi((U^{(s)})^{k}\Psi_{0})=\mu  \; (k =0,1, \ldots, n) \right\}.
\end{align*}
By definition, we see that  
\begin{align*}
{\cal M}_{1} \supseteq {\cal M}_{2} \supseteq  \cdots \supseteq {\cal M}_{n} \supseteq {\cal M}_{n+1} \supseteq \cdots, \quad {\cal M}_s = \bigcap_{n=1}^{\infty} {\cal M}_n.
\end{align*}

As in the case of QWs, we have the following result which is stronger than Theorem \ref{biwako3} by using a similar argument given in \cite{KonnoTakei2014}:
\begin{thm}
For any $U \in \mbox{\boldmath{U}} (2, \HM)$ with $b=0$, we have ${\cal M}_s (U)= {\cal M}_{unif} = {\cal M}_2 (U).$
\end{thm}

\section{Summary \label{sum}}
In this paper, we introduced a QQW determined by a unitary matrix whose component is quaternion and explored the relation between QWs and QQWs. Here we consider the following sets of measures. For a fixed $\varphi \in \Psi (X)$ with $X = \RM, \CM, \HM$, 
\begin{align*}
{\cal M}^{loc} _{n} (\varphi;(U,Y)) 
= \left\{ \phi \left((U^{(s)})^{n} \Psi_{0} ^{\varphi} \right) : U \in \mbox{\boldmath{U}} (2,Y) \right\} \quad (Y = \RM, \CM, \HM).
\end{align*}
For a fixed $\Psi_{0} \in X^{\ZM} \setminus \{ \boldsymbol{0} \}$ with $X = \RM, \CM, \HM$, 
\begin{align*}
{\cal M}^{glob} _{n} (\Psi_{0};(U,Y))
= \left\{ \phi \left( (U^{(s)})^{n} \Psi_{0} \right) : U \in \mbox{\boldmath{U}} (2,Y) \right\} \quad (Y = \RM, \CM, \HM).
\end{align*}
One of the future interesting problems is to clarify the relation among above sets.

\par
\
\par\noindent
{\bf Acknowledgments.} This work was partially supported by the Grant-in-Aid for Scientific Research (C) of Japan Society for the Promotion of Science (Grant No.24540116).

\par
\
\par

\begin{small}
\bibliographystyle{jplain}

\end{small}

\end{document}